# An Inclusive Foundation Model for Generalizable Cytogenetics in Precision Oncology

Changchun Yang[1,2,3,4*], Weiqian Dai[1*], Yilan Zhang[2,3,4*], Siyuan Chen[2,3,4] , Jingdong Hu[5], Junkai Su[5], Yuxuan Chen[5], Ao Xu[5], Na Li[5], Xin Gao[2,3,4#] , Yongguo Yu[1#]

[1]Xinhua Hospital Affiliated to Shanghai Jiao Tong University School of Medicine
[2]Computer Science Program, Computer, Electrical and Mathematical Sciences and Engineering Division, King Abdullah University of Science and Technology (KAUST), Thuwal 23955-6900, Kingdom of Saudi Arabia
[3]Center of Excellence for Smart Health (KCSH), King Abdullah University of Science and Technology (KAUST), Thuwal 23955-6900, Kingdom of Saudi Arabia
[4]Center of Excellence on Generative AI, King Abdullah University of Science and Technology (KAUST), Thuwal 23955-6900, Kingdom of Saudi Arabia
[5]Smiltec(Suzhou)Co.,Ltd
*These authors contributed equally to this work.
#All correspondence should be addressed to YY (yuyongguo@shsmu.edu.cn) and XG (xin.gao@kaust.edu.sa).

## Abstract:

Chromosome analysis is vital for diagnosing genetic disorders and guiding cancer therapy decisions through the identification of somatic clonal aberrations. However, developing an AI model are hindered by the overwhelming complexity and diversity of chromosomal abnormalities, requiring extensive annotation efforts, while automated methods remain task-specific and lack generalizability due to the scarcity of comprehensive datasets spanning diverse resource conditions. Here, we introduce CHROMA, a foundation model for cytogenomics, designed to overcome these challenges by learning generalizable representations of chromosomal abnormalities. Pre-trained on over 84,000 specimens (~4 million chromosomal images) via self-supervised learning, CHROMA outperforms other methods across all types of abnormalities, even when trained on fewer labelled data and more imbalanced datasets. By facilitating comprehensive mapping of instability and clonal leisons across various aberration types, CHROMA offers a scalable and generalizable solution for reliable and automated clinical analysis, reducing the annotation workload for experts and advancing precision oncology through the early detection of rare genomic abnormalities, enabling broad clinical AI applications and making advanced genomic analysis more accessible.

## Main

Karyotype analysis is a foundational assay in medical genetics and precision oncology [1]. By providing a genome-wide snapshot of chromosomal architecture, it underpins the diagnosis of both inherited disorders—such as trisomy 21, trisomy 13/18, Turner syndrome, and balanced translocations responsible for recurrent pregnancy loss—and somatic diseases in which clonal chromosomal aberrations drive malignancy, mostly genetic disorders and cancers [2]. Its ability to detect numerical and structural abnormalities—including translocations, inversions, deletions, duplications, fragments, ring chromosomes, and dicentrics [3]—makes karyotyping the first-line test in prenatal screening, reproductive medicine, and tumor cytogenetics. Despite this broad genetic relevance, conventional karyotyping remains labor-intensive: expert cytogeneticists manually select and review roughly twenty metaphase spreads per case, even when hundreds are available [4]. The throughput bottleneck is compounded in resource-limited settings, where variable staining quality, overlapping chromosomes, and a shortage of trained personnel reduce diagnostic yield. These constraints create a critical need for an automated, scalable, and generalizable karyotype-analysis system that can preserve the diagnostic rigor required for genetic counselling and cancer care while extending high-quality cytogenetic testing to underserved regions, where chromosomal abnormalities are often more prevalent [9–11]. Our study addresses this need by introducing CHROMA, a foundation model that learns universal representations of chromosomal morphology from millions of metaphase images, enabling reliable detection of both germline and somatic abnormalities across diverse clinical scenarios.

To address these challenges and promote accessibility in resource-limited settings, we propose CHROMA, the first foundation model for generalizable cytogenetics, designed to overcome key barriers in automated karyotype analysis. CHROMA tackles three pivotal questions: (1) how to mitigate the effects of poor image quality and overlapping chromosomes, crucial for distinguishing true chromosomal abnormalities from imaging artifacts, especially in areas with limited resources; (2) how to create a robust, universal model capable of detecting a wide range of chromosomal abnormalities, including both numerical and structural changes, especially in complex cases with sparse and imbalanced data; and (3) how to provide a dependable diagnostic tool to obtain trustworthy predictions to minimize misdiagnosis, offering a clinical tool for better understanding the underlying genomic complexities. CHROMA leverages self-supervised pretraining on around 4 million metaphase images and single-chromosome segments (Fig. 1a), encompassing a wide range of real-world variability. A band-guided masking strategy tailored to chromosome topology, combined with noise injection and denoising operations (Fig. 1d), enhances the model's ability to handle common artifacts such as overlapping chromosomes, low resolution, and staining inconsistencies (Fig. 1c). By incorporating a risk-control strategy to differentiate genuine patient-specific aberrations from imaging artifacts, CHROMA significantly reduces false positives, a critical advancement in scenarios where only a limited number of metaphase cells are analyzed per sample.

A key strength of CHROMA is its scalability and adaptability for various cytogenetic tasks. By pretraining on chromosomal patterns, the model can be efficiently fine-tuned for different types of abnormalities, from numerical changes to complex structural aberrations. This capability is particularly beneficial for genetic disorders and hematologic diseases, where rare chromosomal events can profoundly influence diagnosis and treatment decisions. Even under suboptimal imaging conditions, CHROMA can reliably detect and localize these abnormalities, making it a valuable tool across diverse clinical environments.

Despite the widespread challenges in deploying reliable medical AI systems in real-world settings, CHROMA presents a practical solution through two key innovations: a foundation model architecture that captures comprehensive chromosomal patterns, and an extensible risk-control strategy that ensures deployment safety. By automatically identifying cases requiring expert review and maintaining robust safeguards for uncertain predictions, CHROMA serves as a dependable diagnostic support system that enhances both accuracy and workflow efficiency. This capability is particularly valuable in regions with limited access to experienced specialists, where reliable automated support can significantly improve healthcare delivery. Through its interpretable outputs and risk-aware predictions, CHROMA effectively bridges the gap between advanced AI technology and clinical requirements, demonstrating a pathway toward responsible AI deployment in healthcare settings. The system not only advances diagnostic workflows but also establishes a practical framework for integrating AI methodologies into clinical practice while maintaining high standards of reliability and safety.

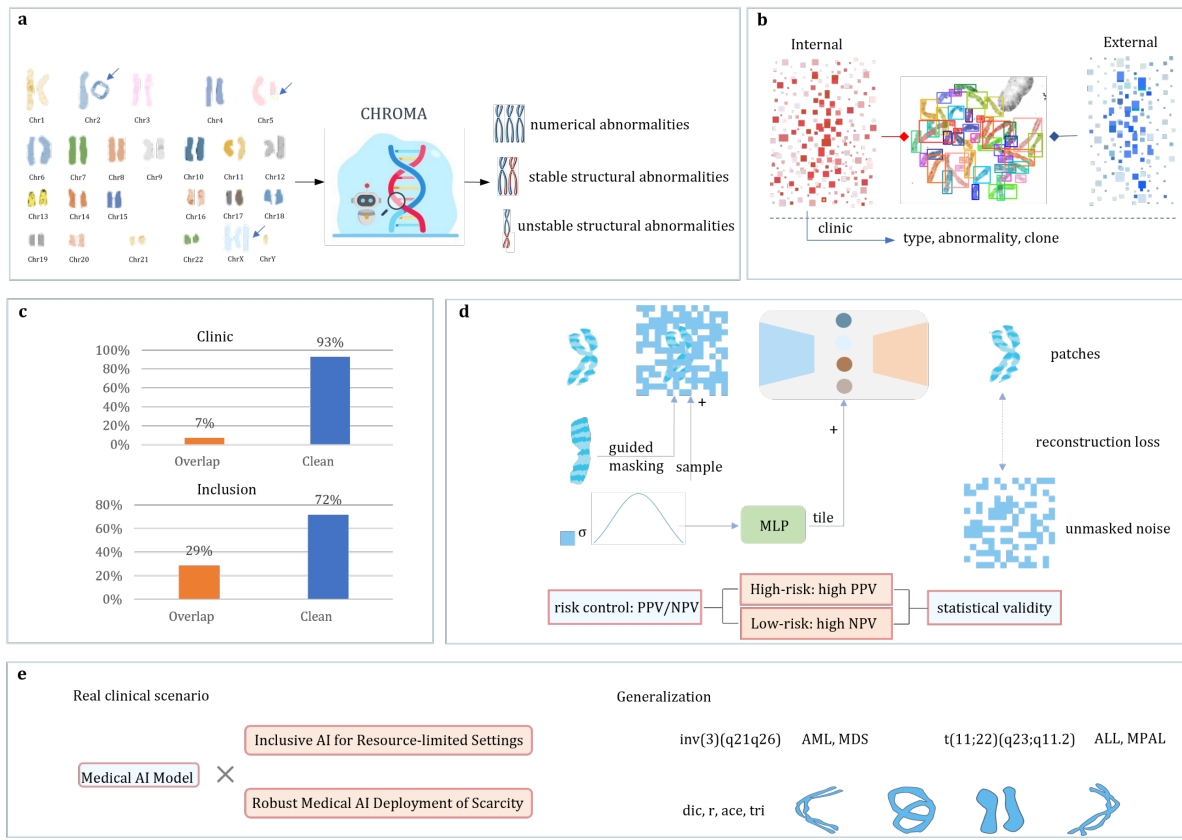

**Fig. 1: Workflow of the CHROMA method and experimental design of the entire study. a** Schematic illustration of the development and evaluation of CHROMA. CHROMA is primarily designed to perform large-scale pretraining on single-chromosome data, which integrates results derived partially across various clinics and partially from metaphase-segmented datasets. Then it can be deployed for diverse clinical applications, including the detection of numerical abnormalities, stability-related abnormalities, and instability-related abnormalities. **b** Data utilization mechanism. Given the rarity of abnormal data, a "human-in-the-loop" strategy was employed for data annotation. **c** Due to variations in sample preparation and process maturity across clinics, the overlap ratio and noise in chromosome data from different sources remain relatively high. This phenomenon can impact abnormality diagnosis, particularly under resource-constrained conditions. **d** Overall pretraining strategy of CHROMA. Based on the masked autoencoder framework, we implemented a band-guided masking strategy tailored for chromosomes. Additionally, we introduced noise-injection and denoising strategies for the unmasked regions. For deployment safety, a risk-control mechanism is integrated into the final prediction stage, where the model either outputs confident predictions or automatically rejects uncertain cases for expert review, ensuring reliable clinical application. **e** Practical application scenarios of CHROMA. Our model robustly identifies abnormalities, particularly for rapid screening in large populations. It can also provide interpretability by revealing relationships between detected abnormalities.

This approach has the potential to advance diagnostic workflows, bridge AI methodologies with cytogenomic needs, and improve patient care.

## Results

### Workflow and datasets

Figure 1 provides an overview of the CHROMA workflow and dataset construction. To build CHROMA, we curated a large-scale dataset consisting of 84,471 specimens, representing approximately 4 million chromosomes. This diverse dataset formed the backbone of CHROMA's pretraining phase. For downstream tasks, we constructed three specialized datasets: (1) a classification dataset with 830,000 chromosomes, annotated using a human-in-the-loop strategy, to systematically study numerical abnormalities such as monosomy and trisomy; (2) a stability abnormality dataset with 50,000 chromosomes, capturing a wide range of structural abnormalities including translocations, inversions, and duplications; and (3) an instability abnormality dataset of the same size, including 1,706 rare abnormalities such as fragments, ring chromosomes, and dicentric chromosomes, to analyze instability-related variations and clonal changes.

The pretraining process used a masked autoencoder [6] framework, enhanced with a band-guided masking strategy tailored to chromosome banding patterns [7], allowing the model to focus on small-scale chromosomal features critical for detecting subtle aberrations. Additionally, we implemented a noise injection and denoising module to address low-quality imaging artifacts, improving robustness under diverse imaging conditions (Fig. 1d). The pretraining phase demonstrated strong alignment with data scaling laws [8] (Extended Data Fig. 1, $R^2 = 0.89$), showing consistent performance gains as data size increased.

After pretraining, CHROMA was fine-tuned using labeled data for specific tasks and evaluated on held-out internal test sets. To benchmark its performance, we compared CHROMA against three baseline approaches: SL, which uses direct supervised learning or transfer learning; SSL-cl, which uses contrastive learning [9] for pretraining; and SSL-MAE, a standard masked autoencoder [6]. All methods were fine-tuned using the same strategy for downstream tasks. Through the integrated risk-control strategy, CHROMA maintains high reliability by only providing predictions for cases meeting strict confidence thresholds, while automatically flagging uncertain cases for specialist review. This mechanism is especially crucial for managing rare or novel chromosomal abnormalities in clinical practice, effectively mitigating the risks associated with imbalanced data distribution in real-world settings. P values were calculated using a two-sided t-test to compare CHROMA with the most competitive baseline model for each task, assessing statistical significance. Details on dataset construction, pretraining, and fine-tuning methodologies can be found in the ***Methods*** section. This approach highlights CHROMA's scalability and adaptability for diverse cytogenetic challenges, particularly in scenarios requiring the detection of highly complex chromosomal abnormalities.

## Identification and numerical abnormalities

To evaluate the performance of CHROMA in chromosome identification and its ability to detect numerical abnormalities, we first examined the distribution and nature of errors using the optimal CHROMA model. The confusion matrix of predicted versus true labels (Fig. 2a) reveals that most discordant predictions occurred between chromosomes of similar size or morphology, such as chr21/chr22. These misclassifications are consistent with challenges faced by human practitioners, who might confuse these chromosomes due to their visual similarities. However, the model-specific error patterns, such as chr15/Y swaps, suggest other biases unique to CHROMA. Notably, discordant predictions were significantly reduced ($P < 0.001$) compared to the most competitive baseline model (Extended Data Fig. 2a). This highlights CHROMA's ability to refine predictions under challenging conditions, achieving an average specificity of 99.8% and sensitivity of 94.9%. To further understand the relationship between concordant and discordant predictions, we visualized the penultimate layer's embeddings of the CHROMA model using UMAP (Fig. 2b). Chromosomes were generally well-separated into distinct clusters, reflecting the model's ability to learn discriminative features. However, discordant predictions were enriched at the periphery of clusters, indicating that the model associates these predictions with less canonical chromosome representations. This is particularly pronounced in cases affected by noise, such as overlapping chromosome images, as shown in the baseline comparison (Extended Data Fig. 2a). In addition to its classification accuracy, CHROMA demonstrated robustness to noise and data imbalance, as shown in Fig. 2c. Across multiple evaluation metrics—including specificity, sensitivity, accuracy, F1-score, MCC, and AUC-ROC—CHROMA consistently outperformed baseline models (85% as basic). This robustness is critical for real-world applications where data quality can vary significantly.

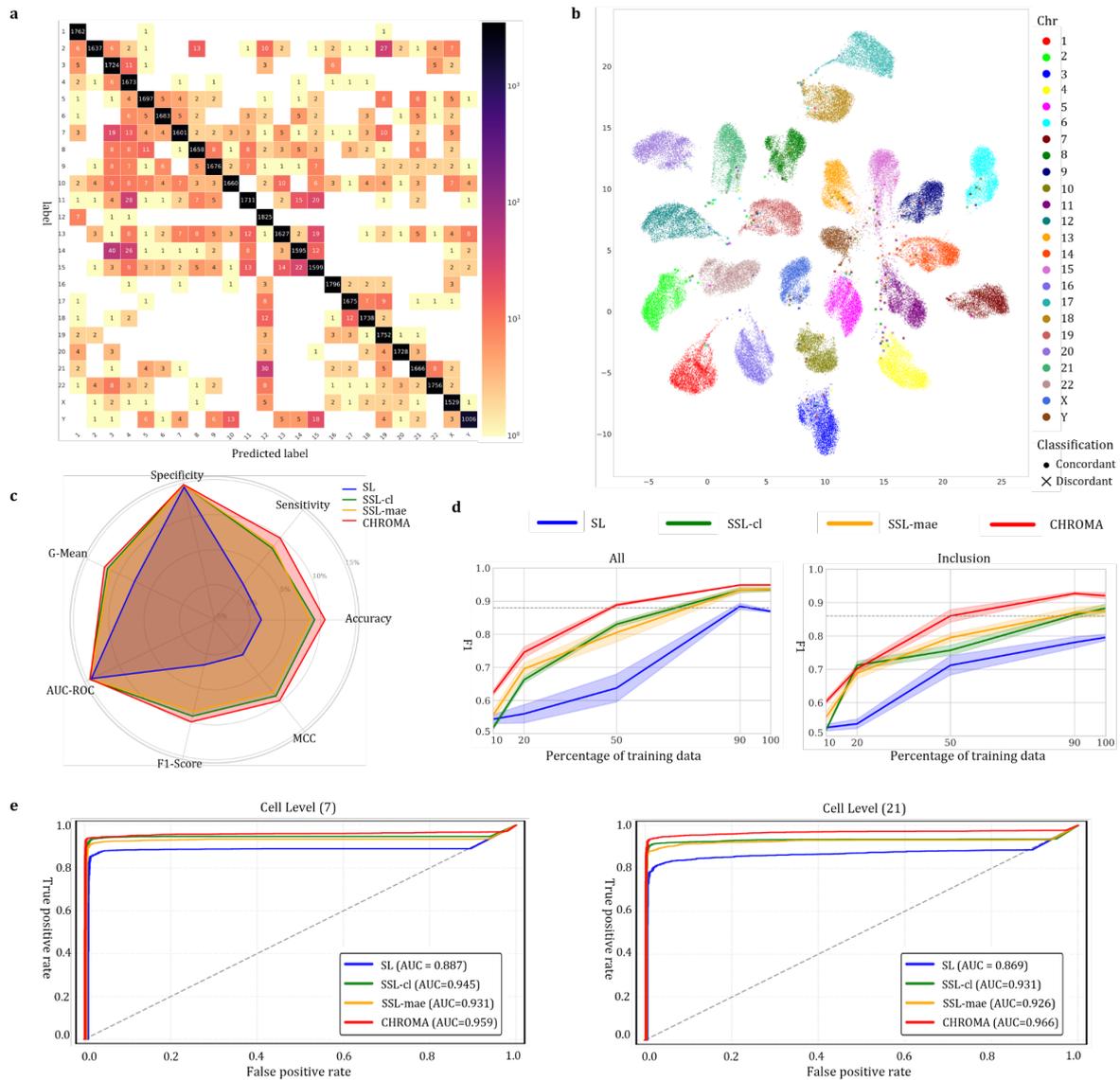

**Fig. 2: Analysis and performance on chromosome identification and numerical abnormalities. (a)** Confusion matrix on the 24-class chromosome classification test set after fine-tuning. CHROMA demonstrates robustness even under conditions of low data quality or class imbalance, achieving an average specificity of 99.8% and sensitivity of 94.9% (P < 0.001). **(b)** UMAP projection of the penultimate layer (prior to the logits layer) of the CHROMA model for test set chromosomes. Each point is colored according to its ground truth label, with misclassified predictions enlarged 3-fold and marked with an X. The clear separation of features highlights the model's precise feature representation. **(c)** Radar chart comparing CHROMA with other methods (SL, SSL-cl, SSL-mae) across multiple performance metrics, including specificity, sensitivity, accuracy, F1-score, MCC, AUC-ROC, and G-Mean. The baseline value for all metrics is set to 0.85. CHROMA (red) consistently outperforms the other methods, demonstrating superior performance across all evaluated metrics. **(d)** Label efficiency in chromosome identification. Label efficiency evaluates performance using varying portions of training data to determine the amount of data required to achieve target performance levels. The dashed gray line highlights the difference in training data requirements between CHROMA and the most competitive baseline model. CHROMA achieves even higher training efficiency when finetuning on Inclusion data, showcasing its robustness against noisy data. The 95% confidence intervals (CIs) for F1 scores are plotted as colored bands, with the center points representing the mean F1 scores. **(e)** ROC curves for chromosomes 7 and 21 at the cell level. These plots evaluate the performance of CHROMA and baseline models (SL, SSL-cl, SSL-mae) in identifying chromosomes 7 (left) and 21 (right). CHROMA achieves the highest area under the curve, AUC = 0.959 for chromosome 7 (P = 0.003) and AUC = 0.966 for chromosome 21 (P < 0.001). The cell-level analysis provides enhanced sensitivity for detecting clinically significant genetic abnormalities, such as monosomy or trisomy, which are critical for diagnosing chromosomal disorders and assessing genetic clonal variations in a clinical setting.

We also assessed CHROMA's label efficiency, which refers to the amount of training data required to achieve a target performance level. CHROMA demonstrated superior label efficiency compared to competing methods (Fig. 2d). For example, in the "Inclusion" dataset, CHROMA achieved an F1-score of 0.86 while using approximately 35% less training data than the most competitive baseline. This efficiency translates to significant reductions in annotation workload and computational costs, making CHROMA highly practical for clinical applications. Given the clinical importance of numerical abnormalities, we conducted a cell-level analysis of chromosomes 7 and 21, which are critical in diagnosing genetic disorders. Monosomy 7 is a hallmark of myelodysplastic syndromes (MDS) and acute myeloid leukemia (AML), while trisomy 21 is strongly associated with acute lymphoblastic leukemia (ALL). CHROMA achieved the highest AUC for both chromosome 7 (AUC = 0.959) and chromosome 21 (AUC = 0.966) (Fig. 2e). This enhanced sensitivity under stricter definitions of abnormality demonstrates CHROMA's utility in identifying key genetic markers of disease.

## Stable structural abnormalities

We assessed CHROMA's performance on stable structural abnormalities by dividing the 24 chromosomes into three categories—low, medium, and high scarcity—based on the number of abnormal samples available (the specific data is distributed in the Extended Data Table. 1). As shown in Fig. 3(a), CHROMA not only achieves superior results across seven common performance metrics (e.g., accuracy, sensitivity, specificity, G-mean, F1-Score AUC-ROC, AUC-PR, MCC) but also exhibits minimal performance degradation when transitioning from low to high scarcity conditions. In contrast, the three baseline methods all experience a decline exceeding 20% in the high-scarcity group, whereas CHROMA maintains an overall performance above 80% ($p < 0.001$). This highlights the robustness of our approach under imbalanced and limited-data scenarios.

In Fig. 3(b), we further illustrate this capability by comparing AUROC values for representative chromosomes chosen from each scarcity group. Even with very few abnormal samples, CHROMA consistently surpasses alternative approaches ($p < 0.001$), indicating that it can effectively adapt to heterogeneous data distributions often found in clinical applications. The AUPR comparison across these three categories is presented in Extended Data Fig. 3.

To delve deeper into the learned representations, Fig. 3(c) provides a UMAP visualization of chromosome 5 in a binary classification context (normal vs. abnormal). Despite being trained solely on a binary objective, CHROMA's embedding space naturally segregates various abnormal subtypes into distinct clusters. In addition to the clear separation of normal cases, each abnormal subtype also forms cohesive patterns—an outcome quantifiable by a k-means clustering purity of 81.8%. This demonstrates CHROMA's inherent capacity to encode fine-grained structural differences, suggesting that only a small number of abnormal samples would be needed to fine-tune the model for newly encountered subtypes. Building on this analysis, Extended Data Fig. 4 provides two additional UMAP visualizations for other chromosomes, further validating the generalizability of CHROMA's representational learning. These visualizations reveal similarly distinct clustering patterns, where abnormal subtypes are cohesively grouped despite the binary training objective. Across these chromosomes, CHROMA consistently demonstrates its ability to learn meaningful latent representations that align with the underlying structural variations. The insights from these UMAP visualizations collectively highlight CHROMA's versatility and representational strength. By effectively distinguishing structural abnormalities and clustering subtypes, CHROMA holds promise for applications in precision medicine where detailed subtype classification and adaptation to novel abnormalities are crucial.

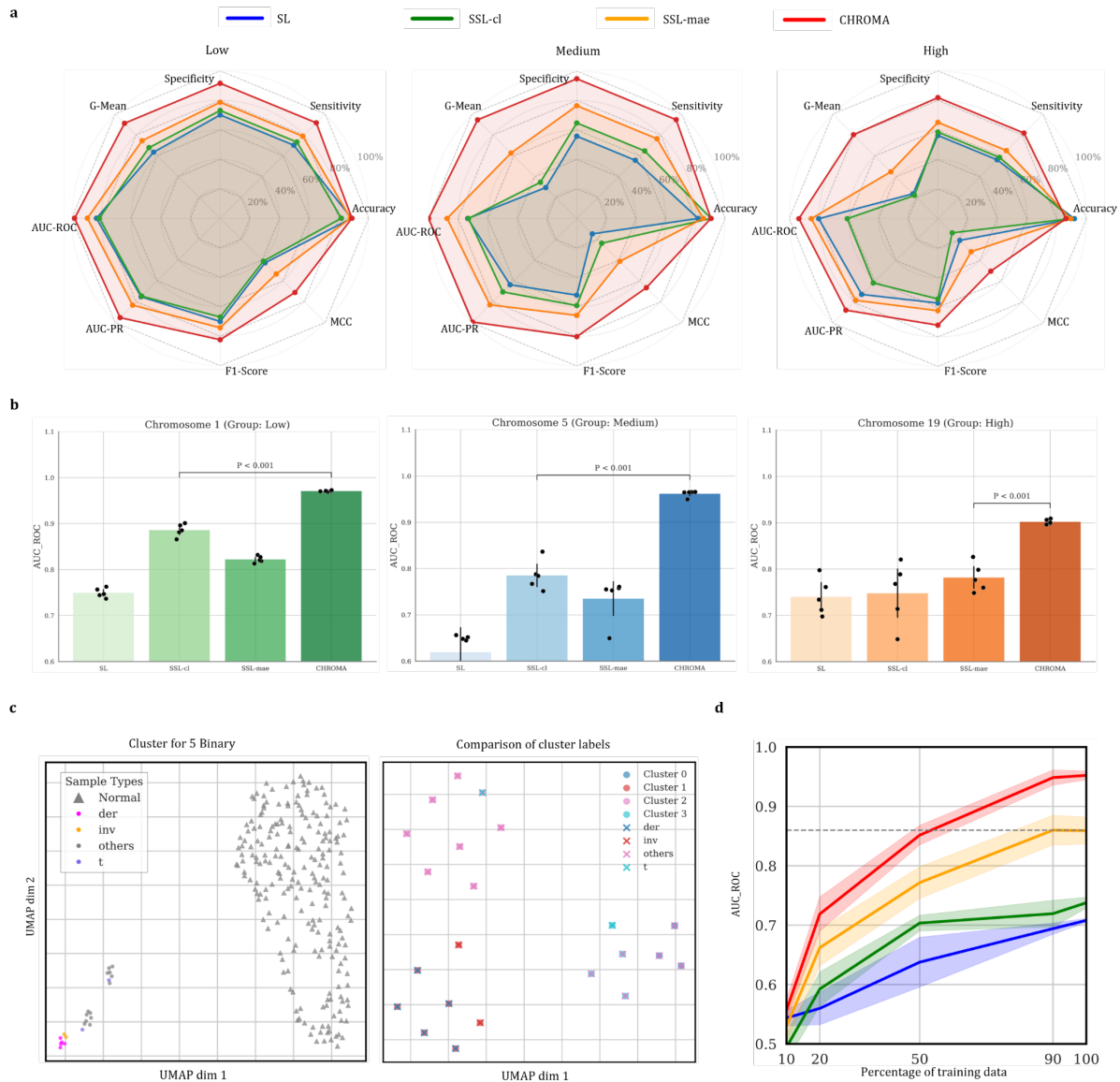

**Fig. 3: Analysis and performance on stable structural abnormalities.** (a) Chromosomes are stratified into three groups (low-, medium-, and high-scarcity) based on the available abnormal data. Despite the imbalanced data distribution, CHROMA consistently outperforms other methods and exhibits markedly lower performance degradation (p < 0.001). (b) AUROC values across all chromosomes demonstrate that CHROMA preserves strong discriminative power even for rare abnormalities (p < 0.001). As examples illustrating the consistent pattern observed across the genome, we highlight improvements of ~8%, ~16%, and ~12% for chromosomes 1, 5, and 19, respectively. These chromosomes were selected to represent different sizes and genomic characteristics. For each class, the model was trained using five distinct random seeds, which determined the shuffling of the training data. The trained models were then evaluated on the test set to generate five independent replicates. Statistical metrics were calculated based on these replicates. The error bars in the plots represent the 95% confidence intervals (CIs), while the bar heights correspond to the mean AUROC values. To assess whether there are statistically significant differences between RETFound and the most competitive baseline model, a two-sided t-test was conducted, and the resulting p-values are reported in the figure. (c) UMAP visualization of chromosome 5 under a binary classification setup (normal vs. abnormal). Although trained purely for binary discrimination, the learned embedding space spontaneously separates different abnormal subtypes (der, inv, others, t) into distinct clusters (k-means purity = 81.8%). This illustrates that the model attains high intra-class coherence and can be readily fine-tuned for newly emerging abnormal classes using only a few additional samples. (d) Label-efficiency analysis shows that CHROMA requires approximately 45% fewer annotated abnormal instances to maintain an AUROC of ~0.86, again highlighting the method's adaptability and practical utility in clinical contexts with limited annotation resources.

Finally, Fig. 3(d) sheds light on the label-efficiency of our method in the context of stable structural abnormalities. CHROMA can reach an AUROC of approximately 0.86 while requiring about 45% fewer labeled abnormal instances than conventional approaches. This substantially reduced annotation burden is especially advantageous in clinical settings, where collecting expertly labeled, high-quality abnormal samples is both time-consuming and expensive, further underscoring the practical feasibility of CHROMA for real-world deployment.

## Unstable structural abnormalities

We explored the performance of CHROMA's on unstable structural abnormalities by evaluating its effectiveness in two separate experimental settings: binary classification (normal vs. abnormal) and five-class classification of abnormal subtypes (ace, dic, min, r, tri). This two-stage approach was essential due to the highly imbalanced data distribution, with a ratio exceeding 1600 between normal and certain abnormal classes (tri). As shown in Fig. 4(a), CHROMA demonstrated superior performance over baseline methods (SL, SSL-cl, SSL-mae) in most of the standard evaluation metrics, including G-mean, AUC-PR, and Sensitivity, while achieving comparable or better results in accuracy, F1-Score, and MCC, in both the binary and five-class settings. This means CHROMA maintained robust performance across all abnormality types, these results highlight CHROMA's resilience to extreme class imbalances and its capability to handle heterogeneous data distributions. Fig. 4(b) presents CHROMA's performance in terms of AUROC values across both binary and five-class classification tasks. The model evaluation was conducted using five distinct random seeds to ensure statistical reliability, with error bars indicating 95% confidence intervals (CIs) and bar heights representing mean AUROC values. In the binary classification task (left), CHROMA achieved significantly higher AUROC ($P < 0.001$) compared to baseline methods, demonstrating its superior capability in distinguishing between normal and abnormal cases. Similarly, in the more challenging five-class classification task (right), CHROMA maintained its leading performance ($P = 0.155$) over other approaches, effectively differentiating among various types of abnormalities. This consistent excellence across both tasks underscores CHROMA's robust and comprehensive diagnostic capabilities in both coarse-grained and fine-grained classification scenarios.

Fig. 4(c) presents confusion matrices comparing the classification performance across three scenarios: SSL-mae, CHROMA, and CHROMA with risk control strategy. In practical clinical applications, both accurate binary detection and precise subtype classification are crucial. CHROMA demonstrated superior performance over SSL-mae, as evidenced by higher diagonal values across all abnormality subtypes, particularly for challenging rare categories like 'min' and 'r'. The implementation of the risk-control strategy further refined CHROMA's performance, leading to a more clinically reliable confusion matrix. Notably, this strategy effectively handles cases involving multiple centromeres, such as 'tri' and 'dic' categories, where confident automated prediction is challenging due to limited training data and structural complexity. These cases are automatically flagged for specialist review, ensuring careful examination of these rare but clinically significant chromosomal patterns. The rightmost column in all matrices reveals CHROMA's more reliable normal/abnormal separation, with the risk-control version showing the cleanest separation, where acc on abnormal classes increased from 0.928 to 0.997 after control.

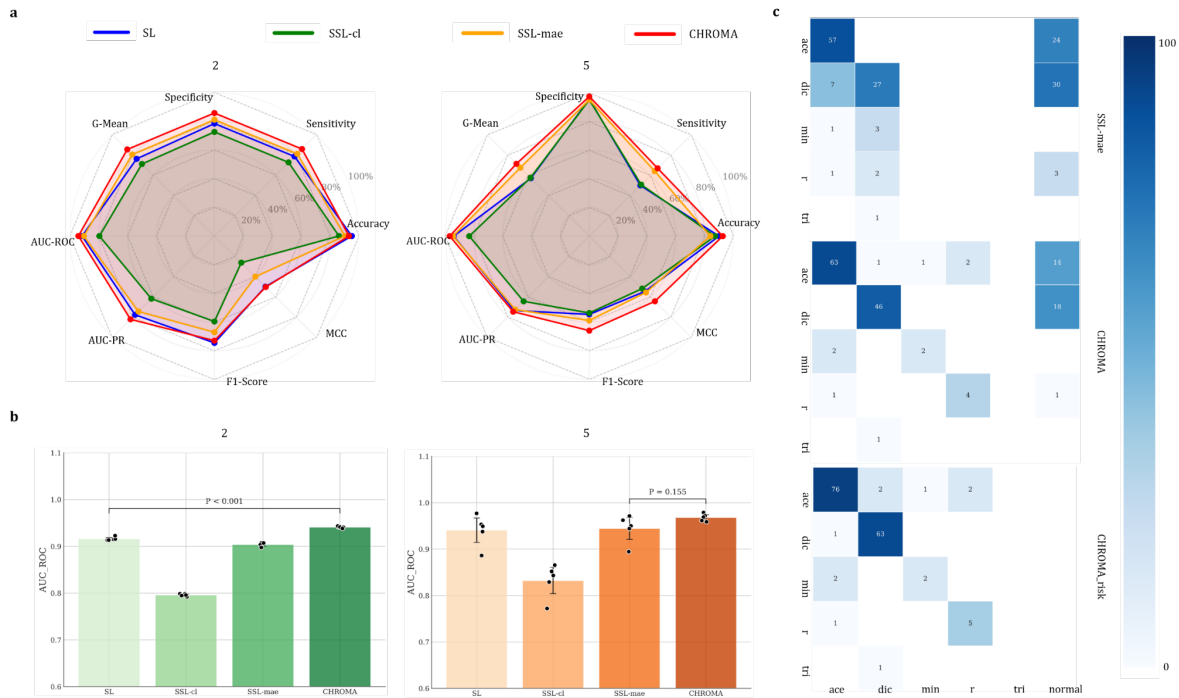

**Fig. 4: Analysis and performance on unstable structural abnormalities. (a)** Comparison of CHROMA and other methods across various performance metrics, including accuaracy, sensitivity, specificity, G-Mean, AUC-ROC, MCC, F1-Score and AUC-PR, for two separate experiments: binary classification (normal vs. abnormal) and five-class classification of abnormalities (ace, dic, min, r, tri). These experiments were conducted independently to account for the significant data imbalance (>1600). **(b)** Representative AUROC values for both binary and five-class classification (full results available in Supplementary), demonstrating that CHROMA maintains strong discriminative power for each abnormality type. Each model was trained with five random seeds, ensuring robust evaluation through shuffled training data and test set replicates. The error bars indicate 95% confidence intervals (CIs), while the bar heights represent the mean AUROC values. A two-sided t-test was performed to evaluate statistical significance, and p-values are shown in the figure, highlighting CHROMA's superior performance. **(c)** In practical applications, final classification performance should integrate both binary and five-class classification accuracies. The confusion matrices show that CHROMA outperforms SSL-mae in detecting various abnormality types, including rare categories like "min" and "r." For chromosomal patterns involving multiple centromeres, such as the extremely rare "tri" category and "dic" category, we flags these cases for specialist review, acknowledging their clinical significance and structural similarity.

This progressive improvement across the three matrices - from SSL-mae to CHROMA to CHROMA with risk control - illustrates how each component contributes to building a more robust and clinically applicable system, effectively balancing high detection sensitivity with reliable subtype classification while appropriately managing uncertainty in extremely rare cases.

# Discussion

In this study, we present CHROMA, to our knowledge, the first inclusive foundation model for cytogenomics that addresses critical challenges in automated chromosome analysis through innovative self-supervised learning and risk-control strategies. Our comprehensive evaluation demonstrates CHROMA's superior performance across various chromosomal abnormalities, from common numerical variations to rare structural aberrations, while maintaining clinical reliability through intelligent risk management. CHROMA's robust performance in handling chromosomal abnormalities, particularly in resource-limited settings, represents a significant advancement in automated karyotype analysis. The model's ability to maintain high accuracy even with reduced training data (requiring approximately 45% fewer labeled samples) while adapting to various imaging conditions addresses a crucial need in clinical cytogenetics. This efficiency is particularly valuable in scenarios where expert annotation resources are scarce, making advanced genomic analysis more accessible to broader populations.

The topology-guided masking strategy and noise injection mechanisms enable CHROMA to effectively handle common technical variations in sample preparation and imaging. This robustness is especially crucial for detecting subtle structural abnormalities, where imaging artifacts could potentially mask or mimic genuine chromosomal changes. By incorporating a sophisticated risk-control framework, CHROMA automatically identifies cases requiring specialist review, particularly for rare abnormalities like tri- and dic-chromosomes, ensuring reliable clinical deployment while maintaining high diagnostic standards.

Moving forward, several aspects warrant further exploration: (1) Integration of additional genomic modalities could enhance CHROMA's diagnostic capabilities, particularly for complex cases where traditional karyotyping alone may be insufficient. (2) The current risk-control strategy could be extended to incorporate dynamic thresholding based on clinical context and specimen quality. (3) Development of interpretability tools could provide deeper insights into the model's decision-making process, particularly for rare abnormalities where clinical evidence is limited. CHROMA's ability to learn comprehensive chromosomal patterns while maintaining deployment safety through risk-aware predictions represents a significant step toward reliable AI integration in clinical cytogenetics. The system's demonstrated effectiveness in handling both common and rare abnormalities, combined with its efficient use of training data, positions it as a valuable tool for advancing precision medicine through improved chromosomal analysis.

Beyond current applications, CHROMA's framework could be extended to analyze more complex genomic arrangements and clonal evolution patterns in cancer progression. The model's ability to capture subtle structural variations while maintaining clinical reliability makes it particularly valuable for longitudinal studies of chromosomal instability in cancer development and treatment response monitoring. These advancements collectively demonstrate how CHROMA addresses the pressing need for reliable, accessible, and efficient chromosomal analysis tools in clinical settings. As genomic medicine continues to evolve, CHROMA's framework provides a foundation for developing more sophisticated AI-driven approaches to chromosomal analysis, potentially enabling earlier detection of genetic abnormalities and more precise therapeutic interventions.

# Data Availability

The anonymized partial data data that support the findings of this study are attached publicly with the trained models. Public datasets in training data for BioImLab (https://www.kaggle.com/datasets/arifmpthesis/bioimlab-chromosome-data-set-for-classification)[12], Pki-3 (https://www.fim.uni-passau.de/en/research-and-professorships/former-chairs-professorships/mathematical-stochastics/chromosome-image-data)[13], CIR-Net (https://github.com/CloudDataLab/CIR-Net/tree/master/data)[14], ChromosomeNet (https://github.com/CloudDataLab/BenchmarkForChromosomeClassification)[15], TVG_Hospital (https://www.cellimagelibrary.org/pages/auto_chromosome_detector)[16], AutoKary2022 (https://github.com/wangjuncongyu/chromosome-instance-segmentation-dataset?tab=readme-ov-file)[17] and CRCN-NE (https://zenodo.org/records/3229434)[18] are publicly available from their original publications. The authors declare that all other data supporting the findings of this study are available within the paper and its supplementary information files.